# Ions leaving no tracks

Azat Abdullaev[1,2], Javier Garcia Fernandez[3], Chloé Nozais[4], Jacques O'Connell[5], Rustem Tlegenov[1,2], Kairolla Sekerbayev[1,2], Alexander Azarov[3], Aleksi Leino[4], Tomás Fernández Bouvier[4], Junlei Zhao[6], Aldo Artímez Peña[7], Nikita Medvedev[7,8], Zhandos Utegulov[2*], Øystein Prytz[3], Flyura Djurabekova[4], Andrej Kuznetsov[3*]

[1]Centre for Energy and Materials Science, National Laboratory of Astana, 010000 Astana, Kazakhstan
[2]Department of Physics, School of Sciences and Humanities, Nazarbayev University, 010000 Astana, Kazakhstan
[3]Department of Physics and Centre for Materials Science and Nanotechnology, University of Oslo, N-0316 Oslo, Norway
[4]Department of Physics and Helsinki Institute of Physics, University of Helsinki, P.O. Box 43, FI-00014, Helsinki, Finland
[5]Centre for HRTEM, Nelson Mandela University, Port Elizabeth, 6001, South Africa
[6]Department of Electronic and Electrical Engineering, Southern University of Science and Technology, Shenzhen, 518055, China
[7]Institute of Physics, Czech Academy of Sciences, Na Slovance 2, 182 21, Prague 8, Czech Republic
[8]Institute of Plasma Physics, Czech Academy of Sciences, Za Slovankou 3, 182 00, Prague 8, Czech Republic

Corresponding authors: andrej.kuznetsov@fys.uio.no; zhutegulov@nu.edu.kz

## Abstract

The paths of swift heavy ions are typically traceable in solids, because of confined electronic interactions along the paths, inducing what is known in literature as "ion tracks", i.e. nano-sized in cross-section cylindrical zones of modified material extending for microns in length. Such tracks readily form in materials exhibiting low thermal conductivities, in particular insulators or semiconductors, altering the homogeneity of materials. In this work, using recently discovered γ/β-Ga$_2$O$_3$ polymorph heterostructures we show that, in contrast to the trends in many other materials, including that in β-Ga$_2$O$_3$, swift heavy ions leave no tracks in γ-Ga$_2$O$_3$. We explained this trend in terms of amazingly fast disorder recovery, occurring because of multiple configurations in the γ-Ga$_2$O$_3$ lattice itself, so that the disorder formed by ion impacts gets rapidly erased, giving a perception of ions leaving no tracks. As such, γ-Ga$_2$O$_3$, readily integrated with β-Ga$_2$O$_3$ in polymorph heterostructures, may become a promising semiconductor platform for devices capable to operate in extremely harsh radiation environments.

## Introduction

Swift heavy ions (SHIs) accelerated to the MeV/u energy range, leave distinct tracks along their paths in solids, because of confined electronic interactions along the paths, rather than stochastically knocking out atoms into blurry collisions cascades, characteristic for slowly accelerated ions[1]. In insulators and semiconductors, SHIs impacts results in what is known in the literature as "ion tracks", *i.e.*, nano-sized in cross-section cylindrical zones of modified material extending for microns in length[2–4]. Such ion tracks can be amorphous[4–6] or crystalline[5,7,8] depending on the SHIs energies and complexity of the materials. In the first approximation, the ion track formation in "complex" and "simple" materials follow different trends because of the varied efficiency of the radiation damage recovery and recrystallization processes, altogether affecting the ion track formation[1,2,4,9]. In this sense, systems with multiple polymorphs exhibiting different thermodynamic stability are very interesting to study. One practically important example of such systems is gallium oxide ($Ga_2O_3$)[10,11], crystallizing into at least five polymorphs known in the literature: α-, β-, δ-, γ-, and ε/κ-phases[12–14]. Moreover, SHIs impacts in $Ga_2O_3$ would be very interesting to study in comparison with the data on its polymorph transitions induced by ions in the keV energy range[15–19] and, specifically, in the context of the outstanding radiation tolerance observed in γ-$Ga_2O_3$ fabricated via disorder-induced ordering[16,20]. Thus, taking into account that the discovery of materials tolerating radiation is among the greatest endeavors in the field, it is particularly interesting to learn whether γ-$Ga_2O_3$ is going to be equally high radiation tolerant against SHIs impacts as it has shown to be for irradiations with keV ions[17].

In this work, we used integrated γ/β-$Ga_2O_3$ polymorph heterostructures[21–23] enabling a direct comparison of the SHIs impacts simultaneously in these two polymorphs. We showed that indeed, in contrast to the trends in many other materials, including that in β-$Ga_2O_3$[24,25], SHIs (at least up to a stopping power of 20 keV/nm) leave no tracks in γ-$Ga_2O_3$, confirming its outstanding radiation tolerance. Notably, γ-$Ga_2O_3$ exhibits the lowest thermal conductivity among the polymorphs in the $Ga_2O_3$ system[21], so that the absence of ion tracks in γ-$Ga_2O_3$ is paradoxical if correlating these phenomena with the heat transfer rates, as it is often done in literature[1,2]. Instead, based on the results of the theoretical modelling, we explained the trends occurring in γ-$Ga_2O_3$ in terms of its intrinsically disordered nature, allowing amazing disorder recovery because of multiple configurations in the γ-$Ga_2O_3$ lattice itself, so that the disorder formed by the SHIs impacts gets rapidly erased, giving a perception of ions leaving no tracks. As such, we conclude that γ-$Ga_2O_3$, can be potentially employed as structural material for nuclear energy applications and can be readily integrated with β-$Ga_2O_3$ by the disorder induced ordering, as a promising semiconductor platform for devices capable of operating in extremely harsh radiation environments.

## Results and discussion

Figure 1 captures a set of scanning transmission electron microscopy (STEM) data revealing that SHIs leave no tracks in the γ-part of the γ/β-$Ga_2O_3$ polymorph heterostructures. Specifically, Fig.

1(a) shows a low magnification cross-sectional image of the sample irradiated with 147 MeV Kr$^+$ ions using a fluence of $5 \times 10^{11}$ ions/cm$^2$. As seen from Fig. 1(a), there are clearly distinguished ion tracks observed in form of the different contrast region repetitions in the β-phase part of the sample. Evidently, the density of these repetitions correlates with the expectations of the impact density given by the fluence. However, there are no corresponding features in the γ-phase, see Fig. 1(a). Such behavior is striking already because of the full chemical identity of the materials around the polymorph interface. Moreover, this trend is generic at least, up to the energies studied in the present work, since we observe similar behavior in the rest of our samples with variable irradiation conditions, see Supplementary note 2.

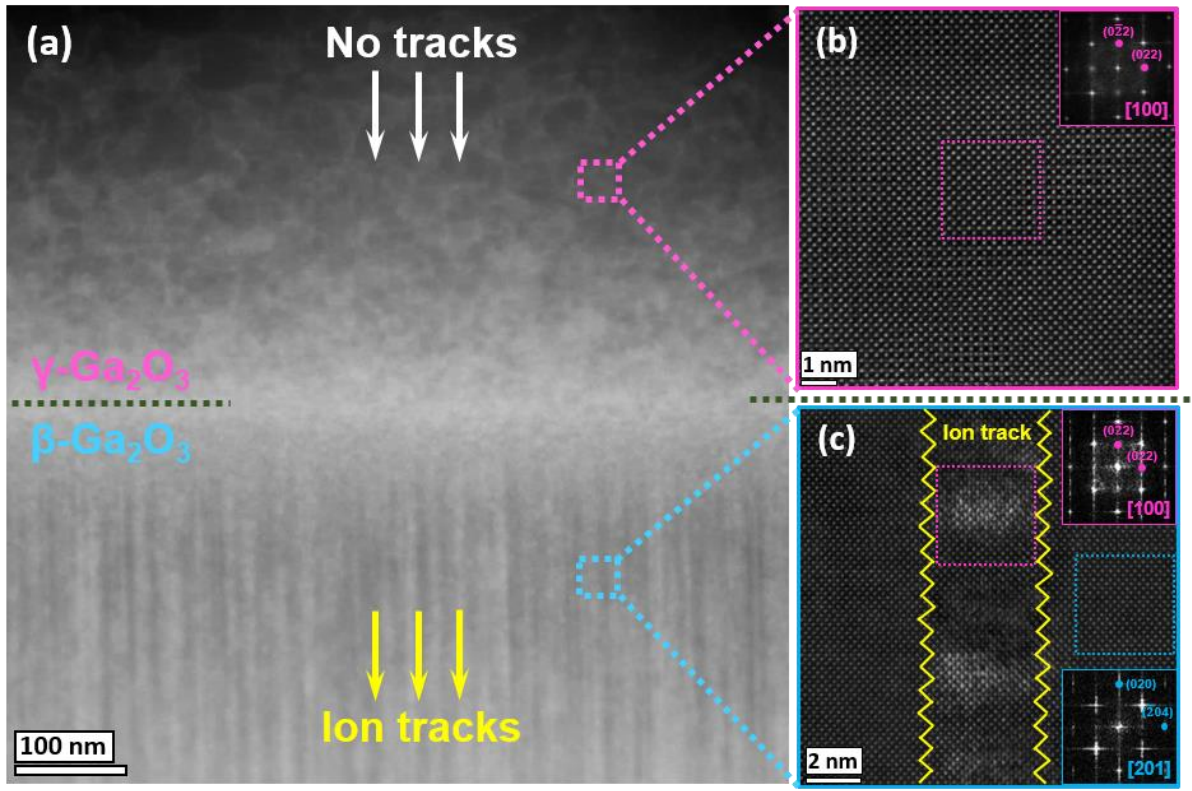

**Figure 1. Scanning Transmission Electron Microscopy (STEM) data revealing that SHIs leave no tracks in the γ-part of the γ/β-Ga₂O₃ polymorph heterostructure**. (a) Low magnification STEM image of the γ/β-Ga₂O₃ polymorph heterostructure irradiated with 147 MeV Kr$^+$ ions using a fluence of $5 \times 10^{11}$ ions/cm$^2$, with arrows indicating the direction of the ion impacts; (b) and (c) atomic resolution STEM images corresponding to the data captured on both sides of the heterostructure with insets representing reciprocal lattice fingerprints recorder by the fast Fourier transforms (FFTs) in corresponding places in the sample (as color-coded).

Importantly, further information is gained by comparing the atomic resolution STEM data, including the corresponding reciprocal lattice fingerprints recorder by the fast Fourier transforms (FFTs), for γ- and β-parts of the sample in Figs.1(b) and 1(c), respectively. As seen from these data, the upper-lying γ-phase part remains structurally unchanged in the form of its defective cubic spinel lattice, without traces of extended defects in Fig.1(b). In contrast, Fig.1(c) confirms the formation of tracks, 3-4 nm in diameter, in the lower-lying β-Ga₂O₃ as measured in a cross-sectional view. Remarkably, these tracks are clearly distinguishable, but they are not amorphous as seen in many other crystalline materials, such as GaN[24] or alpha-quartz[25]. Instead, they are crystalline, but rearranged from the originally monoclinic β-Ga$_2$O$_3$ to the cubic spinel γ-Ga$_2$O$_3$, as confirmed by the FFTs in the inserts in Fig.1(c); *i.e.,* we observed β-to-γ phase transformation inside the tracks in the β-part of the sample in Fig.1. Notably, β-to-γ transformation induced by the keV-accelerated ions has recently been reported in literature as well[15–19,21]. However, the mechanism of the SHI-induced β-to-γ transformation might be different from the purely kinetic pathway, because the energy of the SHIs dissipates via electronic interactions along the paths leading to a transient melting/amorphous state, rather than via a kinetic process of stochastically knocking out atoms from their lattice sites in collisions cascades, which is characteristic of the keV energy range ions[1]. From a thermodynamic perspective, γ-Ga$_2$O$_3$ can be considered as an energetically competitive disordered form of β-Ga$_2$O$_3$ at high temperatures[26] further facilitated by the anisotropic tensile strain and/or localized Ga deficiency[27]. Consequently, the SHI-induced heterogeneous β → melting/amorphous → γ transformation pathway within the initial β-Ga$_2$O$_3$ matrix involves both thermodynamic (at high temperatures) and kinetical (at low temperatures) driving forces.

Nevertheless, already from Fig. 1, intuitively, trends of (*i*) 147 MeV Kr$^+$ ions leaving no tracks in γ-Ga$_2$O$_3$ and (*ii*) β-to-γ phase transformation inside the tracks in β-Ga$_2$O$_3$ might be interconnected. Moreover, since the SHIs track phenomena are often correlated with the heat transfer rates, we performed depth-resolved thermal conductivity measurements across our γ/β-Ga$_2$O$_3$ polymorph heterostructure samples, as shown in Fig.2. Indeed, Fig.2(a) is a summary of the data plotting the normalized thermal conductivity as a function of the Kr$^+$ ion energy as measured in the γ/β-Ga$_2$O$_3$ polymorph heterostructures in the pristine ($k_0$) and upon SHIs irradiation ($k_{irr}$) states. The data reveals two clear trends. Firstly, there is a drop in the normalized thermal conductivity ($k_{irr}/k_0$) in β-Ga$_2$O$_3$ as a function of the ion energy in Fig.2(a), correlated with a higher damage fraction with increasing ion energy[28,29]. Secondly, in γ-Ga$_2$O$_3$, $k_{irr}/k_0$ remains nearly constant, correlating with a hypothesis of the maintained homogeneity of the material, consistent with data in Fig.1, see eye-guiding horizontal line in Fig.2(a). To see a more detailed picture, Fig.2(b) provides an example of the depth-resolved thermal conductivity profile through the γ/β-Ga$_2$O$_3$ polymorph heterostructure picking up the data for the sample irradiated with 50 MeV Kr$^+$ ions, in comparison with the data for the pristine sample illustrated by cartoons in the lower and upper parts of the panel, respectively. It is clearly seen from Fig.2(b) that in the near-surface region, coinciding with the thickness of the top γ-Ga$_2$O$_3$ film (see "no tracks" label), there is practically no thermal conductivity change before and after irradiation – confirming maintained homogeneity of the γ-Ga$_2$O$_3$ film. In contrast, deeper in the sample, beyond the γ/β-Ga$_2$O$_3$ interface (see "tracks" label), there is a clear deviation between the data before and after irradiation. Moreover, the deviation is in the form of the thermal conductivity decrease which is consistent with the fact of increasing γ-Ga$_2$O$_3$ fractions embedded into the β-Ga$_2$O$_3$ matrix (*i.e.,* with tracks observed in Fig. 1 as well as Figs.S1-S3 shown in SI). Thus, we conclude that the data in Figs.1 and 2 are in excellent agreement. Notably, more detailed explanations of the thermal conductivity measurements and additional measurements data are available from the Supplementary note 3.

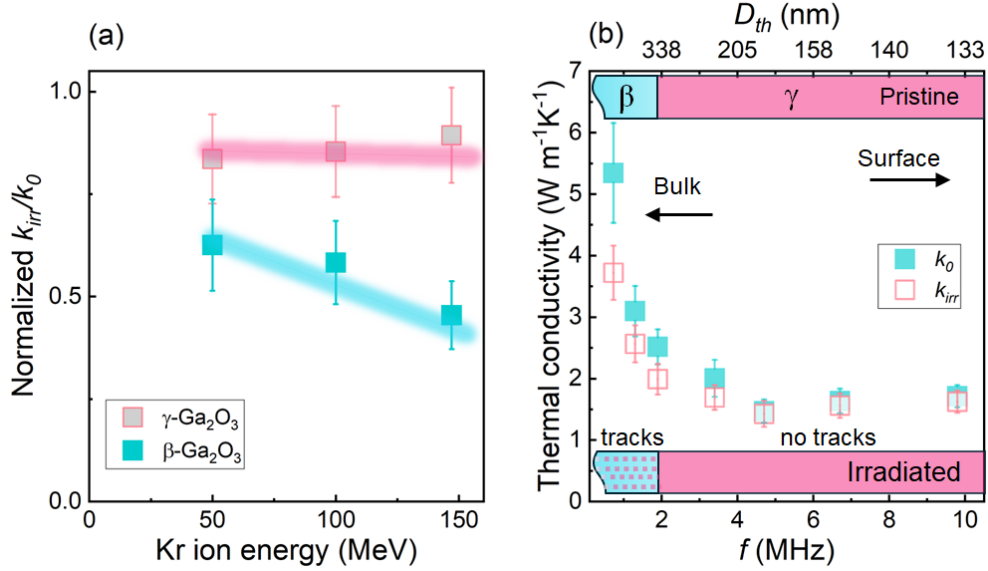

**Figure 2. Thermal conductivity data measured by the time-domain thermoreflectance (TDTR) and frequency domain thermoreflectance (FDTR) through the γ/β-Ga₂O₃ polymorph heterostructures in their pristine state ($k_0$) and upon SHIs irradiations ($k_{irr}$).** (a) Normalized $k_{irr}/k_0$ data plotted as a function of the ion energy with eye-guiding lines highlighting two different trends for γ- and β-parts of the samples. (b) Depth-resolved thermal conductivity profiles measured by TDTR comparing the pristine state with the changes induced by the 50 MeV Kr⁺ ion irradiation, with schematics of the sample inserted for clarity. All data in this figure are collected for 5×10¹¹ Kr/cm² irradiation fluence, enabling direct comparison with data in Fig.1.

Further, in accordance with data in Fig.2, γ-Ga₂O₃ exhibits lower thermal conductivity compared to that in the β-phase, so that no ion tracks in γ-Ga₂O₃ is somewhat counterintuitive, if simply correlating these phenomena with the heat transfer rates.[1] Indeed, if tracks form in β-Ga₂O₃ they might form in γ-Ga₂O₃ too, as it consists of exactly the same atoms – if displacement energies matter – and having even lower thermal conductivity, translated into its lower ability to dissipate energy transferred in SHIs impacts, allegedly[1,2].

To understand this paradox, we employed a hybrid/multiscale model, capturing both the electronic system kinetics after irradiation and the resulting energy transfer to the lattice, altogether translated into the atomic response. Importantly, we reveal that during femtosecond timescales, corresponding to the post-irradiation electron cascades, both polymorphs exhibit nearly identical behavior (see **SI-4.1.1.1-2** for more detail), while divergence emerges during longer timescales when atomic motion and further relaxation takes place, as summarized in Fig.3. Here we show two sets of snapshots of the irradiated β-Ga₂O₃ and γ-Ga₂O₃ simulation cells in Figs.3(b) and (c), respectively. The left images in each set show the structures of the tracks formed immediately after the SHIs impact, or more precisely upon 200 ps allowing the kinetic energy of the atoms within the track to dissipate into the bulk. The right-hand side snapshots show the track upon long-term evolution (4 ns at 1250 K, see the Methods section for details). To visualize the disorder within the tracks, we color-coded O atoms. Accounting that O atoms follow the face-centered cubic (FCC) stacking in both β and γ-phases, it is instructive to color code O atoms in the FCC and non-FCC positions differently, namely in red and blue in Fig.3, allowing quantification of the disorder within the tracks. Indeed, the blue-colored O atoms bands observed immediately (200 ps) after ion impacts in both polymorphs reveal formation of the ion tracks, i.e. amorphous regions, of 4.7 ± 0.5 and 4.3 ± 0.2 nm in diameter in the β- and γ-phases, respectively. Amorphization of the lattice by SHIs impact is expected, as it has been seen in many materials having predominantly covalent bonding[24,25]. At the same time, in some materials with dominating ionic bonding, the SHIs induced amorphization may be avoided[9]. Notably, Ga$_2$O$_3$ exhibits a mixture of ionic and covalent bonding. Hence it is not trivial to predict the nature of ion tracks in this material, not least accounting for its complex polymorphic nature. Overall, the amorphous tracks formation in both β- and γ-phases immediately after SHIs impacts, see Fig.3, is consistent with the sufficient energy exchange between electrons and atoms, leading to practically simultaneous displacement of all atoms within the track. However, the atomic displacements are shorter at the periphery of the ion track, because of the lower kinetic energies gained by atoms in these regions. The shorter displacements[9] along with interfacing the pristine lattice around the track, enables O-sublattice recovering at the outer ion track borders immediately after the impact, as evidenced by the differently colored contrast around the edges of the remaining amorphous track in the left-hand side panel of Fig.3(b). Importantly, this extra contrast at the outer region represents a newly formed γ-phase with the O-sublattice returned to its FCC stacking, while the heavier Ga atoms were reorganized forming the defective spinel structure, see more details in the Supplementary Videos in the SI. Such amorphous-to-γ recrystallization at the edges of the tracks – immediately after impacts – occurs in the γ-Ga$_2$O$_3$ sample too, effectively reducing the tracks size, see the left-hand side panel in Fig.3(c). As a result, a striking difference in behavior is revealed upon the long-term evolution, i.e. in right-hand side panels in Figs.3(b) and (c). Indeed, ~95 % of the disordered O atoms return to the FCC sites as seen from the matching color codes in the track and out of track regions in Fig.3(c), with only a few residual point defects left. Nearly complete amorphous-to-γ-recrystallization takes place in the β-phase too, resulting in the γ-phase ion tracks there, see the corresponding data in Fig.3(b) upon the long-term evolution with a final track diameter of 5.3 ± 0.3 nm. In contrast, in the γ-phase the tracks are erased, and this recovery persists even for significantly larger initial amorphous regions (see Fig.S14 in the Supplementary Information) indicating a universal recovery mechanism and, supporting the experimental observation that ions leave no tracks in Fig.1. Note that we studied the long-term structural evolution in MD by increasing temperature, effectively lowering energy barriers so that recovery occurs within the simulation timescale. To ensure that simulations at elevated temperatures follow the room temperature evolution of the experiments, we ran simulations over a broad temperature range; the resulting linear trend permits a reliable extrapolation of the recovery time to room temperature (see Fig.S13 in the Supplementary Information). Notably, at temperatures above 1250 K, this linearity breaks, and amorphous tracks in β-Ga$_2$O$_3$ recrystallize back to the initial lattice rather than to γ-Ga$_2$O$_3$.

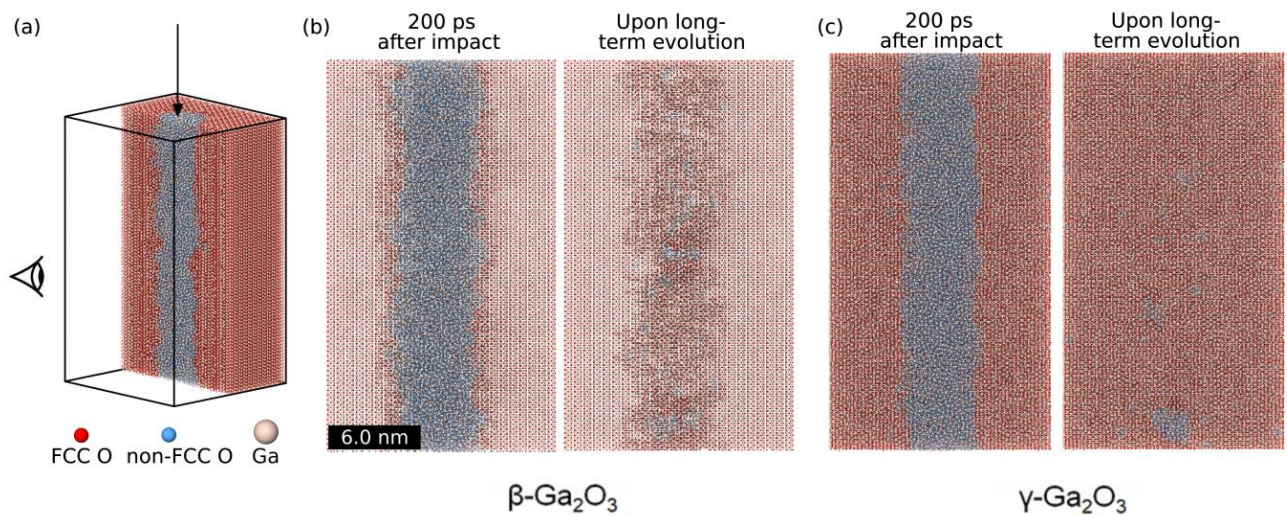

**Figure 3. Summary of the representative molecular dynamic (MD) modelling data for the 147 MeV Kr⁺ ion impacts into β- and γ-Ga₂O₃.** (a) Cross-section of the simulation cell illustrating structural changes in the ion track, visualized by discriminating O atoms in the FCC and non-FCC-like positions. The eye icon indicates the position of the viewer relative to the cell projection shown in panels (b) for β- Ga₂O₃ and (c) for γ-Ga₂O₃. The data in panels (b) and (c) were collected both immediately (200 ps) after the impacts and upon long-term evolutions.

Thus, in accordance with Fig.3, the amorphous-to-γ recrystallization simply erasing the tracks in γ-Ga₂O₃, in excellent agreement with the experimental data and as already mentioned above, giving a perception of ions leaving no tracks in this material. At this end, to verify quantitative agreement between the theoretical and experimental atomistic images, we performed an additional back-loop simulation of the microscopy data, feeding it with the atomic configurations obtained in Fig.3. These data are shown in Fig.4 comparing β- and γ-Ga₂O₃ samples. Notably, the data in Fig.4 immediately (200 ps) after the impact – in panels (a) and (c) – are somewhat illusive, considering its comparison with the data in Fig.1, at least for our experimental conditions with samples kept at room temperature, as such unavoidably experiencing long term evolutions performed theoretically. On contrary, the simulated microscopy images in Figs.4(b) and (d) excellently matching the data in Fig.1, both in terms of the track nature and size – in panel (b) and in terms of indistinguishability for microscopy to capture the residual point defects contents predicted by theory – in panel (d), see side-by-side comparison of the simulated and experimental STEM images in Supplementary note 5.

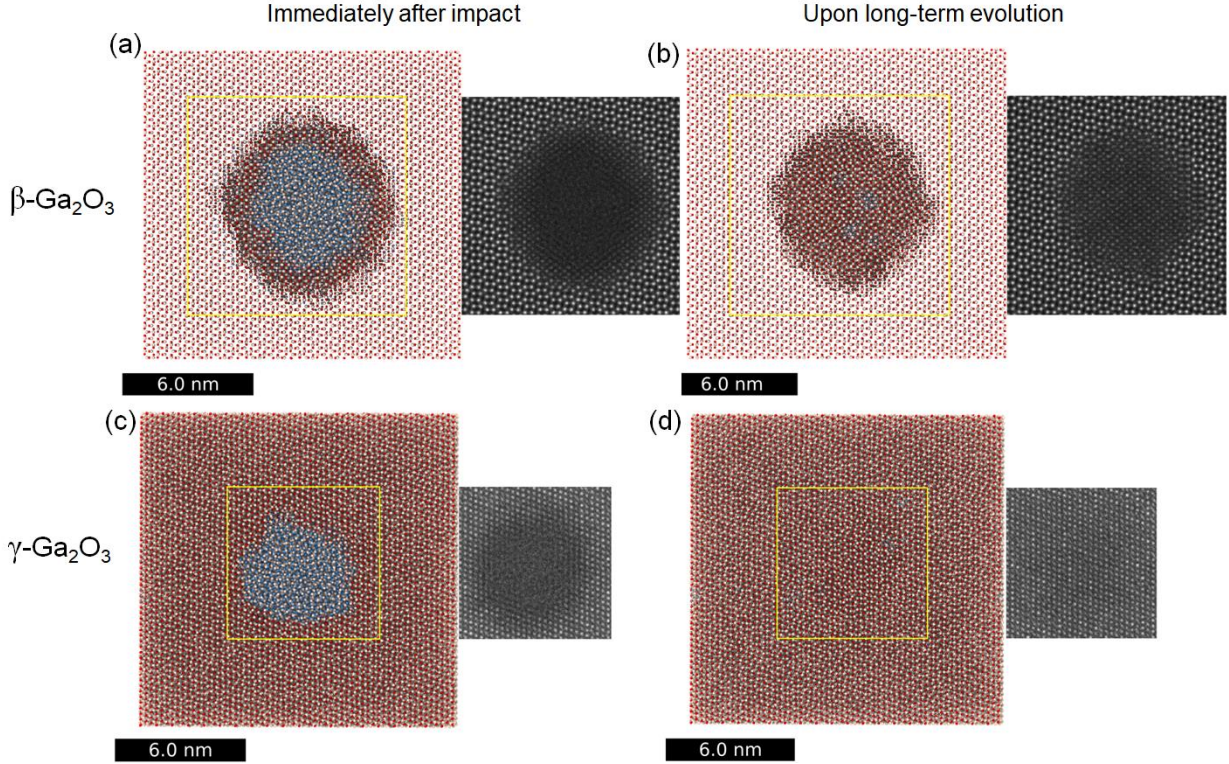

**Figure 4. Results of the back-loop simulation of the microscopy data, feeding it with the atomic configurations obtained with MD**. (a) – (d) pairs of atomistic maps obtained with MD and corresponding STEM images for β- $Ga_2O_3$ and γ-$Ga_2O_3$ immediately (200 ps) after irradiation with 147 MeV $Kr^+$ ion and upon long-term evolutions. In the MD cells, Ga and FCC O atoms are in beige and red, respectively while non-FCC O are in blue. The black boxes in the atomistic maps indicate the boundaries of the STEM simulation. The color coding of the atoms is the same as in Fig.3.

## Conclusions

Combining systematic experimental and modelling approaches, we show that, in contrast to the behaviour in many other materials, including that in β-$Ga_2O_3$, swift heavy ions leave no tracks in γ-$Ga_2O_3$ across a wide range of conditions, at least up to the ion stopping powers of ~20 keV/nm. Importantly, γ-$Ga_2O_3$ exhibits lower thermal conductivity compared to that in the β-phase, so that the absence of ion tracks in γ-$Ga_2O_3$ is somewhat counterintuitive. Indeed, if tracks form in β-$Ga_2O_3$ they might form in γ-$Ga_2O_3$ too, since it exhibits exactly the same FCC packing of the O sublattice and having the same cations. Moreover, lower thermal conductivity in γ-$Ga_2O_3$ translates into lower ability to dissipate energy transferred in SHIs impacts, intuitively provoking track formations. We resolve this paradox by showing that γ-$Ga_2O_3$ undergoes fast disorder recovery, enabled by multiple lattice configurations. As a result, disorder from ion impacts is rapidly erased, giving a perception that ions leave no tracks. This recovery persists even when the initial disorder is larger than expected, confirming extraordinary radiation tolerance of the γ-phase. As such, γ-$Ga_2O_3$, readily integrated with β-$Ga_2O_3$ in polymorph heterostructures, may become a promising semiconductor platform for devices capable to operate in extremely harsh radiation environments.

***Experimental and Computational Methods***

*Samples and irradiations*

$\gamma/\beta$-$Ga_2O_3$ polymorph heterostructures and $\beta$-$Ga_2O_3$ bulk crystals were used in these studies. The heterostructures were fabricated by the disorder-induced ordering[16] of (010) $\beta$-$Ga_2O_3$ single crystal wafers (Novel Crystal Technology Inc.) to have three different $\gamma$-$Ga_2O_3$ film thicknesses: 350, 650 and 1000 nm. In addition, (010) $\beta$-$Ga_2O_3$ single crystal wafers were used for comparison. Both $\gamma/\beta$-$Ga_2O_3$ polymorph heterostructures and $\beta$-$Ga_2O_3$ wafers used for comparison were then irradiated with Kr ions. We used 3 Kr ion energies – 50, 100 and 147 MeV with two ion fluences $1 \times 10^{10}$ and $5 \times 10^{11}$ ion/$cm^2$. Supplementary information provides a full survey of the samples and irradiations used in this study, see Supplementary note-1.

*Electron microscopy*

For Scanning Transmission Electron Microscopy (STEM) studies, selected samples were thinned by mechanical polishing and by Ar ion milling in a Gatan PIPS II (Model 695), followed by plasma cleaning (Fishione Model 1020) immediately before loading the samples into a microscope. STEM imaging was done at 200 kV in a Cs-corrected Thermo Fisher Scientific Titan G2 60–300 kV microscope. The STEM images were recorded using a probe convergence semi-angle of 23 mrad, a nominal camera length of 60 mm using two different detectors: high-angle annular dark field (HAADF) (collection angles 100–200 mrad), and annular dark field (ADF) (collection angles 22–100 mrad). Plan view TEM lamellae were produced by FIB lift out using a Thermo Fischer Helios 650 dual beam. Imaging was done at 200 kV in a Cs corrected JEOL ARM 200F. Images were recorded using a probe convergence angle of 22 mrad and collection angle 62-247 mrad. In order to directly compare the results of MD simulation with experimental STEM images, multislice image simulation was performed with abTEM[30] using the MD supercell as structural input. The interatomic potential was sampled at 0.005 nm in the x and y dimension using Lobato parameterization[31] and slice thickness of 0.2 nm. Due to limited computational resources the MD cell was trimmed to approximately 14x14 nm in x and y limiting the angular extent of the CBED pattern to 167 mrad. Images were simulated for a probe convergence angle of 22 mrad and detector collection angle of 62-167 mrad.

*Thermal conductivity measurements*

For thermal conductivity measurements, two complementary thermoreflectance techniques, specifically time-domain (TDTR) and frequency-domain (FDTR) were employed. To start with, an identical $\sim 90$ nm Al transducer layer was deposited on pristine and irradiated samples via magnetron sputtering. The details of both TDTR and FDTR methods are given extensively elsewhere in literature[28,32,33] and supplementary materials. Briefly, in TDTR measurements a Ti:sapphire femtosecond laser (782 nm, 80 MHz, 80 fs) was used as pump and probe beams, focused to ~11 μm spot size using a 10× objective. The pump, modulated in the MHz range, induced thermal waves, while the time-delayed probe signal ($-V_{in}/V_{out}$) was analyzed using a heat diffusion model[34] to extract the thermal conductivity and thermal boundary conductance between Al transducer and sample. In FDTR setup we used 660 nm and 785 nm CW diode lasers as pump and probe beams, respectively. Unlike TDTR, in FDTR a square-wave modulated pump (in the kHz range) enabled thermal waves to reach deeper into the sample capturing both irradiated and unirradiated regions. Beams were focused to ~1 μm using a 60× objective and dichroic mirror. The phase lag between the pump and probe signals, recorded by a lock-in amplifier, was fitted using a heat diffusion model[35,36] to obtain the thermal conductivity. FDTR was especially suitable for $\beta$-$Ga_2O_3$ bulk crystal samples, as Kr ion range (6–11 μm) spans beyond the TDTR depth sensitivity. More details are available in the Supplementary Information.

*Monte Carlo and Molecular dynamics simulations*

Atomistic simulations are performed using LAMMPS[37] with a machine learning potential (tabGAP) for $Ga_2O_3$ structures[38]. We generate two simulation cells of $\beta$-$Ga_2O_3$ and $\gamma$-$Ga_2O_3$, both 24 × 24 × 25 nm in size. To capture the random order in which Ga atoms partially occupy the octahedral and tetrahedral sites in $\gamma$-$Ga_2O_3$, the Ga atoms in the $\gamma$-phase simulation cell were displaced by 6 Å in a random direction to mimic its spinel structure. The operation was followed by energy minimization to let the structure reach a local energy minimum before performing further simulations. The temperature was raised to 1500 K at 0 GPa in anisotropic isothermal-isobaric ensemble (*NPT*) at the rate of 12 K/ps, followed by the annealing in the canonical ensemble (*NVT*) for 2 ns. Further, the temperature was decreased to 300 K at a rate of 12 K/ps. Both cells were finally relaxed in anisotropic *NPT* at 300 K and 0 GPa for 50 ps to prepare for the impact simulations. To simulate the effect of the 147 MeV Kr ion, instantaneous radial distributions of energy were created with the Monte-Carlo code TREKIS-3[39] for $\beta$-$Ga_2O_3$ and $\gamma$-$Ga_2O_3$, as explained in Supplementary note 4. The simulation cells are oriented such that the ion passes along the longest cell direction, 25 nm, corresponding to [010] for $\beta$-$Ga_2O_3$ and [110] for $\gamma$-$Ga_2O_3$. After depositing the energy, the systems were let to evolve for 200 ps in the microcanonical ensemble (*NVE*) with a variable time step to ensure numerical stability of the simulation. 1 nm of the borders perpendicular to the ion's path are cooled down to 300 K using the Berendsen thermostat with a temperature damping parameter of 0.01 ps to mimic heat conduction to the bulk during the impact simulation. Long-term evolution of the structures is simulated by the subsequent annealing, see more details about the simulated long-term evolution in Supplementary note 4.2.2. For that, the irradiated systems were cut in the x- and y-directions, while preserving periodicity, to speed up the simulation. The resulting size of the simulation cells were 14 × 14 × 25 nm. The temperature was increased to 1250 K in anisotropic *NPT* at 0 GPa at a rate of 12 K/ps (see the main text for the rationale behind this temperature choice). Once the target temperature is reached, the system was kept at in *NVT* for 4 ns. The temperature is brought back down to 300 K by the same method. The visualizations were created with the OVITO PRO software[40] and the Polyhedral Template Matching (PTM)[41] method is used to quantify the damage to the O-sublattice in both phases as they share the FCC structure.

**Data availability**

The data that support the findings of this study are available within the paper and its Supplementary Information file.

**Code availability**

The code and software used in this work are LAMMPS, OVITO, and SRIM, which are openly available online from the corresponding developers and maintainers. The Monte-Carlo code TREKIS-3 used to model SHI impact and electron kinetics is available from https://doi.org/10.5281/zenodo.12704728

## Acknowledgements

M-ERA.NET Program is acknowledged for financial support via the GOFIB project (administrated by the Research Council of Norway project number 337627 in Norway and the Academy of Finland project number 352518 in Finland). Additional support was received from the DIOGO project funded by the Research Council of Norway in the frame of the FRIPRO Program project number 351033. The experimental infrastructures were provided at the Norwegian Micro- and Nano-Fabrication Facility, NorFab, supported by the Research Council of Norway project number 295864, at the Norwegian Center for Transmission Electron Microscopy, NORTEM, supported by the Research Council of Norway project number 197405. Additionally, this work was supported by the Science Committee of the Ministry of Science and Higher Education of the Republic of Kazakhstan [Grants AP19577063, AP19679332], Nazarbayev University grants via Collaborative Research Program (CRP) [111024CRP2003], and Faculty Development Competitive Research Grants Program (FDCRGP) [20122022FD4130] in Kazakhstan. Financial support from the European Commission Horizon MSCA-SE Project MAMBA [HORIZON-MSCA-SE-2022 GAN 101131245] is acknowledged by NM and AAP. Computational resources for TREKIS-3 simulations were provided by the e-INFRA CZ project (ID:90254), supported by the Ministry of Education, Youth and Sports of the Czech Republic. NM thanks the financial support from the Czech Ministry of Education, Youth, and Sports (grant nr. LM2023068).

## Authors contribution

A.K., and Az.A. conceived the research strategy and designed the methodological complementarities. Al.A. prepared γ/β heterostructure samples for this study. Az.A., R. T., and K.S. carried out thermal transport measurements and provided initial description of the data analysis. J.G.F. and J.O. carried out electron microscopy experiments as well as back loop image simulations and provided the description of these data. C. N., A.L., T.F.B., J.Z., and F.D. performed molecular dynamics simulations and provided the description of this data. A.A.P. and N.M. conducted the Monte Carlo simulations and supplied relevant documentation. A.K. and Az.A. finalized the manuscript with inputs from all the co-authors. All co-authors discussed the results as well as reviewed, edited and approved the manuscript. A.K., Ø.P., F.D., J.Z., and Z. U. supervised and administrated their parts of the project and contributed to the funding acquisition. A.K. coordinated the work of the partners.

## Competing interests

The authors declare no competing interest.

# Supplementary information

# Ions leaving no tracks

Azat Abdullaev[1,2], Javier Garcia Fernandez[3], Chloé Nozais[4], Jacques O'Connell[5], Rustem Tlegenov[1,2], Kairolla Sekerbayev[1,2], Alexander Azarov[3], Aleksi Leino[4], Tomás Fernández Bouvier[4], Junlei Zhao[6], Aldo Artímez Peña[7], Nikita Medvedev[7,8], Zhandos Utegulov[2*], Øystein Prytz[3], Flyura Djurabekova[4], Andrej Kuznetsov[3*]

[1]Centre for Energy and Materials Science, National Laboratory of Astana, 010000 Astana, Kazakhstan
[2]Department of Physics, School of Sciences and Humanities, Nazarbayev University, 010000 Astana, Kazakhstan
[3]Department of Physics and Centre for Materials Science and Nanotechnology, University of Oslo, N-0316 Oslo, Norway
[4]Department of Physics and Helsinki Institute of Physics, University of Helsinki, P.O. Box 43, FI-00014, Helsinki, Finland
[5]Centre for HRTEM, Nelson Mandela University, Port Elizabeth, 6001, South Africa
[6]Department of Electronic and Electrical Engineering, Southern University of Science and Technology, Shenzhen, 518055, China
[7]Institute of Physics, Czech Academy of Sciences, Na Slovance 2, 182 21, Prague 8, Czech Republic
[8]Institute of Plasma Physics, Czech Academy of Sciences, Za Slovankou 3, 182 00, Prague 8, Czech Republic

**Corresponding authors*: andrej.kuznetsov@fys.uio.no; zhutegulov@nu.edu.kz**

**TABLE OF CONTENTS**

# Supplementary note 1: Map of the samples, irradiations, and measurements

**Table S1.** Summary of the samples and irradiations used in the present study together with a map of the data included in the manuscript and possible to make it available upon request.

| Samples | Fluence (ion/cm$^2$) | Kr ion energy (MeV) | STEM | TDTR/FDTR | MD | STEM back-loop simulation |
|---------|----------------------|---------------------|------|-----------|-----|----------------------------|
| 1000 nm γ/β | $5\times10^{11}$ | 147 | Fig. 1 | Fig. 2 (a) | Fig. 3 | Fig. 4 |
| 650 nm γ/β | $5\times10^{11}$ | 100 | available | Fig. 2 (a) | available | available |
| 350 nm γ/β | $5\times10^{11}$ | 50 | available | Fig. 2 (a) and (b) | available | available |
| β-(bulk) | $1\times10^{10}$ | 147 | Fig. S3 | available | available | available |
| β-(bulk) | $5\times10^{11}$ | 50 100 147 | Fig. S1 and S2 | Fig. S3 | Fig. 3 | Fig. 4 |

**Supplementary note 2: STEM of ion tracks in β-Ga₂O₃ crystals**

In the main text of the paper, we obviously focused on the prime topic of ions leaving no tracks in γ-Ga₂O₃, while observations of ion track formation in β-Ga₂O₃ had a supportive function: to highlight the paradox of such different behavior in chemically identical materials. On the other hand, in this sub-section we summarize the systematic data on ion track formation in β-Ga₂O₃ as a function of ion energy and fluence.

Figure S1 provides a direct comparison of the track formation phenomena in β-Ga₂O₃ with that in the γ/β-Ga₂O₃ polymorph heterostructure applying identical irradiation conditions, albeit 147 MeV irradiation with 5×10¹¹ Kr⁺/cm² ions. The irradiated surface is located at the top of the image covered with an amorphous carbon layer (indicated) that was deposited during the FIB procedure. Indeed, as seen from Fig.S1(a) – the BF TEM cross-sectional image – the overall structure of the ion tracks in the β-Ga₂O₃ crystal appears to be similar to that in the β-Ga₂O₃ part of the γ/β-Ga₂O₃ polymorph heterostructure in Fig.1(a). Notably, consistently with the ion fluence used for these irradiations, ion tracks in these samples start to overlap, as revealed by the HAADF STEM plan-view image with differently shaped tracks appearing in darker contrast in Fig.S1(b). Individual tracks vary significantly in size. Importantly, Fig.1(c) captures an atomically resolved plan-view image of the track indicated by the white arrow in (b) in the surrounding matrix, including FFTs taken from the areas indicated by the yellow and white boxes, apparently revealing γ-Ga₂O₃ and β-Ga₂O₃ in these color-coded areas, respectively. Moreover, careful investigation of the atomic resolution images, shows that the newly formed γ-Ga₂O₃ in the ion tracks embedded into the β-Ga₂O₃ matrix follow the crystallographic relationship: [100] γ-Ga₂O₃//[201] β-Ga₂O₃ and [01$\bar{1}$] γ-Ga₂O₃//[010] β-Ga₂O₃[1,2], as confirmed by the corresponding FFTs too. In this way, the formation of the γ-Ga₂O₃ ion tracks embedded into the β-Ga₂O₃ matrix is unambiguously confirmed in a direct comparison to the data in Fig.1 significantly strengthening the arguments, because the plan view observations of the tracks in the β-Ga₂O₃ part of the γ/β-Ga₂O₃ polymorph heterostructure was shadowed by the overlaying γ-Ga₂O₃ film, see Fig.1(a).

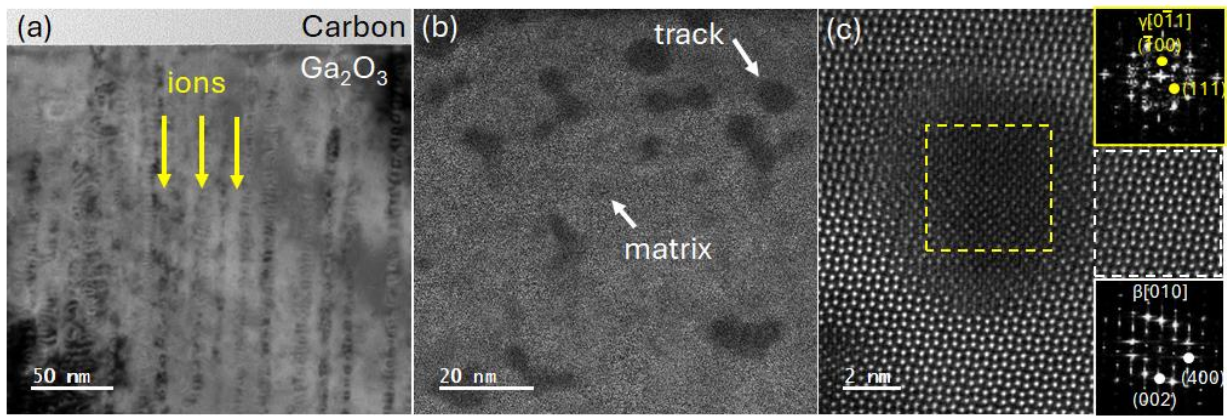

**Figure S1. Electron microscopy data on the formation of ion tracks in β-Ga₂O₃ crystal as a result of 147 MeV irradiation at 5×10¹¹ Kr⁺/cm² fluence, i.e. using identical irradiation conditions as for the γ/β-Ga₂O₃ polymorph heterostructure irradiation in Fig.1.** (a) BF TEM cross-sectional image taken at low magnification; (b) and (c) HAADF STEM plan-view images

at two different magnifications resolving the structure of the tracks. For clarity, panel (c) includes FFT taken from the indicated areas (yellow and white boxes) corresponding to γ-Ga₂O₃ and β-Ga₂O₃, respectively.

Further, to enable a direct comparison of the thermal conductivity depth profiles in Fig.2(b) measured for specimens irradiated with 50 MeV Kr⁺ to 5×10¹¹ ions/cm² fluence, Figure S2 shows BF TEM cross sectional and plan view HAADF STEM images obtained in β-Ga₂O₃ crystal for these irradiation conditions in panels (a) and (b), respectively. An atomically resolved HAADF image of the track marked by the arow in (b) is shown in (c) with FFT insets similar to those in Figure S1(c). Indeed, as seen from Fig.S2(c), the ion tracks in this sample are similar to those observed for the higher energy irradiations in Figs.S1 and Fig.1 and also exhibit the β- to γ-phase transformation.

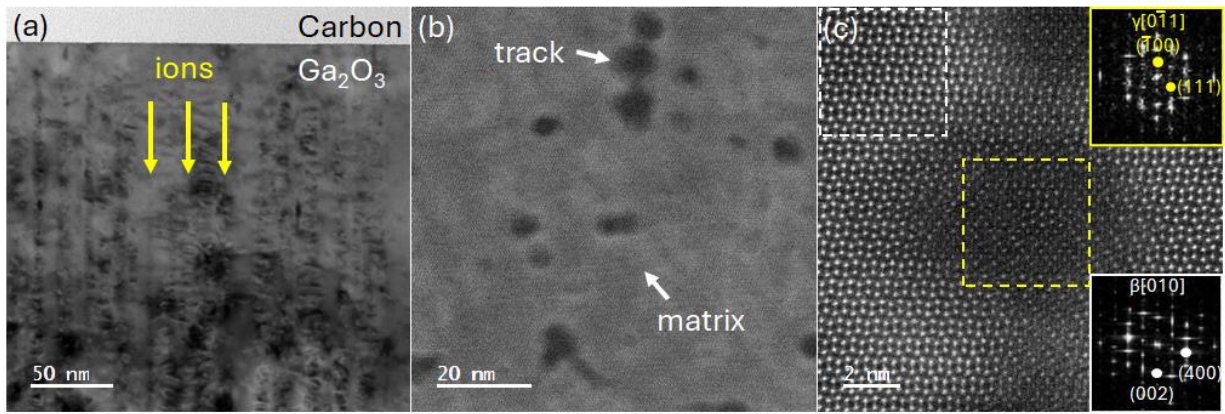

**Figure S2. Electron microscopy data collected from β-Ga₂O₃ crystal irradiated with 50 MeV Kr+ to a fluence of 5×10¹¹ ions/cm², confirming β-to-γ phase transition in ion tracks at this lower ion energy.** (a) BF STEM cross-sectional image taken at low magnification and (b) HAADF STEM plan-view image of several ion tracks with (c) atomically-resolved image with γ-Ga₂O₃ ion tracks resolved in the β-Ga₂O₃ matrix.

Finally, for completeness, we investigated the structure of tracks in samples irradiated with lower fluence, and Figure S3 provides examples of such data for 147 MeV Kr⁺ irradiations of β-Ga₂O₃ crystal with 1×10¹⁰ ions/cm². (a-b) show a low magnification ADF and atomic resolution HAADF STEM cross section of the sample SHIs implanted with low dose such as 1×10¹⁰ Kr⁺ and 147 MeV on β-Ga₂O₃. The bright vertical lines observed in Fig.S3(a) can be identified as individual ion tracks formed by the implantation of the SHIs. Looking in detail in the atomic resolution images of Fig.S3(b) we can observe that the ion track is around 3-4 nm in diameter.

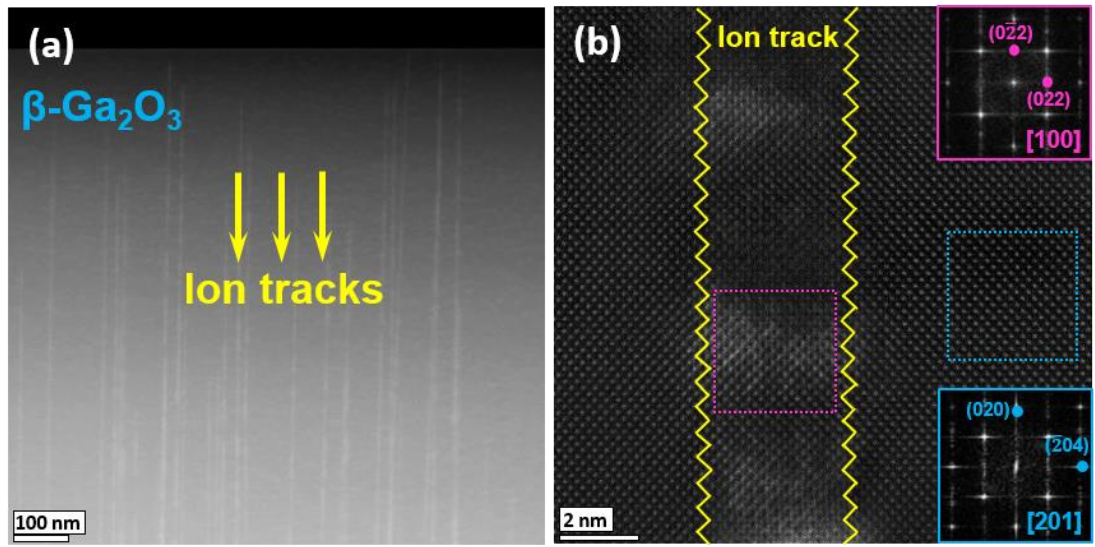

**Figure S3. Electron microscopy data collected from β-Ga₂O₃ crystal irradiated with 147 MeV Kr+ to a fluence of 1×10¹⁰ ions/cm², confirming formation of the individual γ-phase ion tracks in the β-phase matrix at this lower ion fluence.** (a) ADF STEM cross-sectional image at low magnification revealing ion tracks in form of brighter contrast vertical lines and (b) plan view HAADF STEM atomically resolved image FFTs taken from different areas (pink and blue boxes) corresponding to γ-Ga₂O₃ and β-Ga₂O₃, respectively.

**Supplementary note 3: Extended explanations of thermal measurements**

As already stated in the main text of the paper (in the corresponding methodology description), we used two complimentary methods for the thermal conductivity depth profiling, TDTR and FDTR, see measurements schematics in Figs.S4(a) and (b), primarily because of the appropriate combination for the depth sensitivity regions of the two, as illustrated in Figs.S4(c) and (d). In TDTR, the pump beam, modulated by an electro-optic modulator over 0.73-10 MHz thermally excites the samples and control the heat penetration depth by $D_{th} = \sqrt{k/\pi C_v f}$, where $k$ is the thermal conductivity, $C_v$ is the volumetric specific heat and $f$ is the pump modulation frequency. Importantly, for the given modulation frequency range, $D_{th}$ varies from 300 nm to 1000 nm as such limiting TDTR range of the thermal conductivity. To fit the data in the heat diffusion model, we considered a multi-layer structure consisting of a thin transducer layer, an irradiated layer, and a semi-infinite unirradiated region as illustrated in Fig.S4(c) aligned with the $D_{th}$, as well as 50, 100 and 147 MeV penetration depth as plotted in Fig.S4(d) in accordance with TRIM calculations. In this way, TDTR was applicable to study thinner γ/β-Ga₂O₃ polymorph heterostructures.

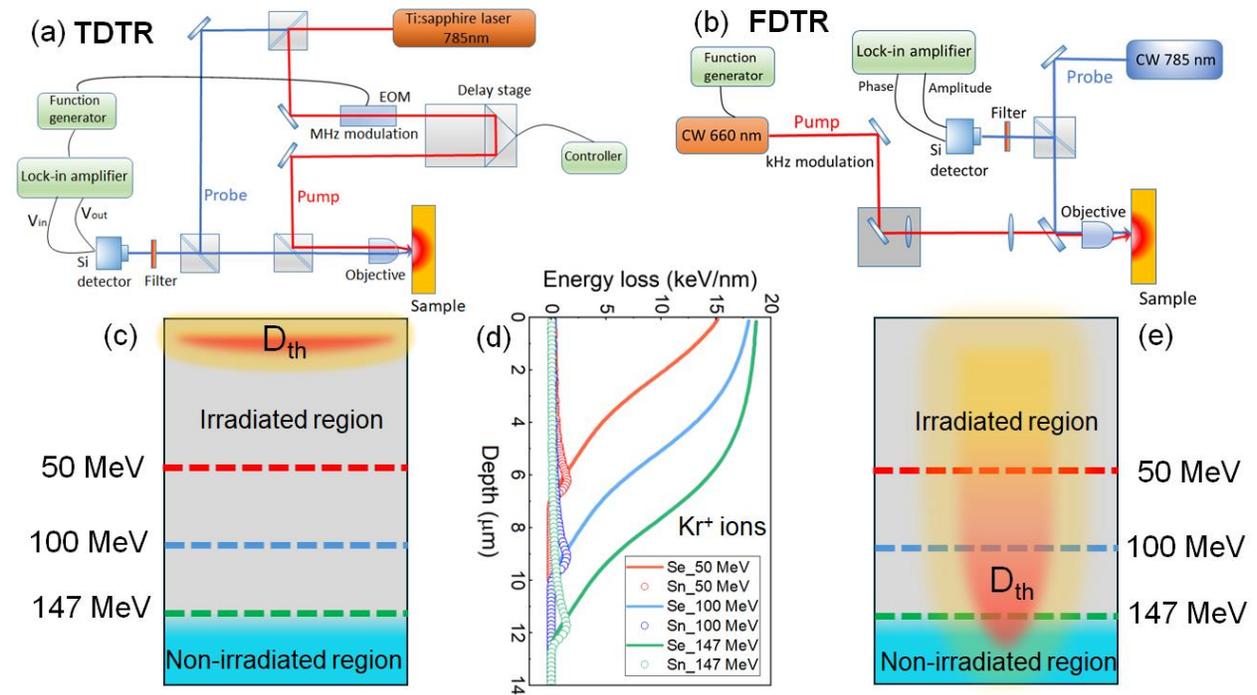

**Figure S4. Thermoreflectance measurements and their depth resolutions.** (a) and (b) Schematics of TDTR and FDTR setups used in this study. (c-e) Cartoons explaining depth sensitivity regions for TDTR (panel (c)) and FDTR (panel (e)) in relation with SRIM data (panel (d)) plotting penetration depth of the Kr⁺ ions depending on the contribution from the electron energy losses (Se-lines) and nuclear energy losses (Sn-symbols).

In contrast, for FDTR the pump modulation in the range of 1 – 100 kHz results in $D_{th} > 30$ μm so that thermal waves penetrate both irradiated and non-irradiated regions of β-Ga₂O₃, see Fig.S4(e). Notable, higher sensitivity in thermoreflectance measurements enables more accurate extraction of thermal conductivity data. For that reason, we performed the sensitivity analysis following the formalism in the previous works and the data are summarized in Figs.S5. As shown in Fig.S5(a), the TDTR sensitivity analysis for the structure in Fig.S4(c) indicates that the measurement is most sensitive to the cross-plane thermal conductivity of the irradiated region, while the non-irradiated region shows negligible sensitivity due to low values of $D_{th}$. Similarly, Fig.S5(b) presents the FDTR sensitivity results for the layered structure in Fig.S4(e), revealing that the measurement is also primarily sensitive to the cross-plane thermal conductivity of the damaged layer. However, owing to the greater heat probing depth in FDTR, there is additional sensitivity to the thermal conductivity of the non-irradiated region, particularly at lower modulation frequencies. Thus, based on this sensitivity analysis in Fig.S5, the total experimental uncertainty was estimated by considering contributions from uncertainties in Al transducer thickness (±3%), irradiated $Ga_2O_3$ layer thickness (±5%), volumetric heat capacity of $Ga_2O_3$ (±5%), and laser spot size (±5%), as e.g. taken into account by the error bars in Fig.2 in the main text.

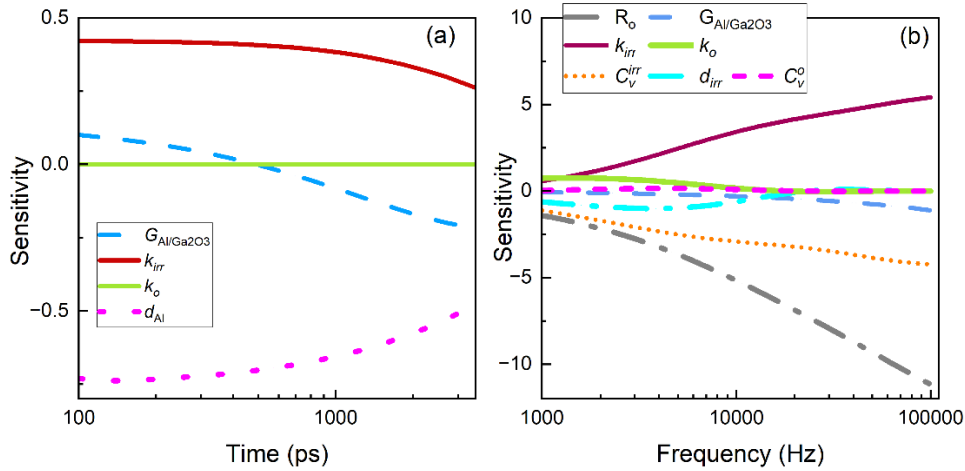

**Figure S5. Sensitivity analysis of 147 MeV irradiated sample for (a) TDTR data and (b) FDTR data.** The notations are the following, $k_o$ and $k_{irr}$ label data for the non-irradiated and irradiated thermal conductivities. G is the thermal boundary conductance between transducer and $Ga_2O_3$, while $d_{Al}$ is the thickness of transducer layer. Additionally in panel (b), $C_v^o$ and $C_v^{irr}$ plot the volumetric specific heat capacity of the pristine and irradiated regions, while $d_{irr}$ and $R_o$ show the thickness of damaged layer and beam spot size.

Figure S6(a) presents the absolute thermal conductivity data extracted from TDTR as a function of the Kr$^+$ ion energy in the γ/β-$Ga_2O_3$ polymorph heterostructures in the pristine and after SHIs irradiation. As a reference to the irradiations of bulk β-$Ga_2O_3$ crystals, Fig.S6(b) summarizes the extracted thermal conductivity values from both TDTR and FDTR data, in relation to the unirradiated β-$Ga_2O_3$ sample exhibiting thermal conductivities exceeding 20 W m⁻¹K⁻¹, in good agreement with reported values for this crystallographic orientation[5]. Upon irradiation, thermal conductivity in both γ- and β-$Ga_2O_3$ decreases, but only in β-$Ga_2O_3$ it decreases significantly and pronouncedly in the highest ion energy irradiated sample, dropping by more than a half of the pristine value, while in γ-$Ga_2O_3$ it remains unchanged with ion energy. The reduction in β-$Ga_2O_3$ is attributed to an increasing fraction of γ-$Ga_2O_3$ ion tracks with increasing ion energy, consistently with STEM analysis, see Figs.S1-S3. These tracks severely disrupt long-wavelength phonon propagation, resulting in substantial thermal conductivity degradation even at relatively low fluences used in this work.

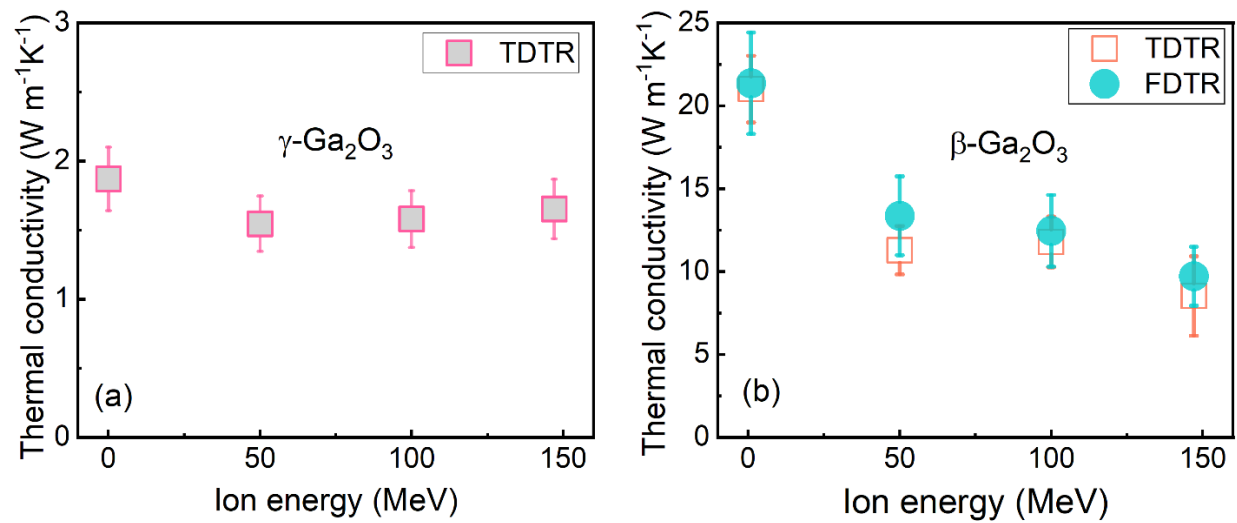

**Figure S6. Summary of extracted thermal conductivity values from TDTR (squares) and FDTR (circles) as a function of the Kr⁺ ion energy**. (a) $\gamma/\beta$-Ga₂O₃ polymorph heterostructures and (b) $\beta$-Ga₂O₃ crystals.

**Supplementary note 4 : Detailed descriptions of computational modelling**

**SI 4.1 Monte Carlo Modelling**

The Monte Carlo (MC) code TREKIS-3 (Time-Resolved Electron Kinetics in SHI-irradiated Solids[6]) is used to describe the impact of 147 MeV Kr ion on the β and γ phases of $Ga_2O_3$[7,8]. TREKIS-3 code relies on the asymptotic trajectory event-by-event MC simulations of individual particle propagation to model the swift heavy ion (SHI) penetration and ionization of the target; the kinetics of δ- and secondary electrons; the Auger and radiative decays of deep shell holes; transport and absorption of the photons produced, as well as the valence holes transport and their interaction with target atoms[9].

In the MC, the target is assumed to be an isotropic and homogeneous arrangement of atoms with densities corresponding to the solid densities of each phase, and accounting for the differences in the average speed of sound and the electronic band gap (Table S2)[9]. The SHI is simulated as a point-like particle with the effective charge calculated with the Barkas formula[9–12]. The free-flight distance of an SHI (as well as of electrons, valence holes, and photons) is sampled using the mean free path calculated with help of the complex dielectric function (CDF) formalism[9,13]. It includes collective responses of the electronic system and the lattice (scattering on plasmons and phonons, respectively)[9,13,14]. The CDF parameters for the ß and γ-phases of $Ga_2O_3$ are derived using the single-pole approximation (see Section 4.1.1. for validation of the cross sections)[15].

**Table S2. Parameters used in the single-pole approximation of CDF.**

| Phase | Density (g/cm³) | Speed of sound (m/s)[1] | Band gap (eV) |
|---|---|---|---|
| $\beta$-$Ga_2O_3$ | 5.88 | 4975[16,17] | 4.84[18] |
| $\gamma$-$Ga_2O_3$ | 5.94[19] | 5041[2] | 5.0[20] |

Electrons are ionized from deep atomic energy levels or the valence band, according to the density of states (DOS) of the material[21,22]. The atomic energy levels (ionization potentials), Auger and radiative decay times of deep-shell holes are taken from EPICS2023 database[23].

The propagation and scattering of excited electrons are simulated accounting for the inelastic (impact ionization) scattering, and scattering on the target lattice (quasi-elastic scattering; electron-phonon interaction). A particular realization of a scattering event — an impact ionization of the valence band or atomic shell electrons versus quasi-elastic scattering on the lattice — are chosen according to the partial cross-sections of these processes.

---

[1] The value is an average across all modes and crystallographic directions.
[2] Calculated.

The spatial propagation of valence holes is simulated similarly to that of electrons, taking into account the hole's effective mass calculated from the DOS of the valence band within the effective one-band approximation[8,24].

All these processes are traced in TREKIS-3 until the end of the electron cascade, which is typically ~100 fs, after which the density of excited electrons in the center of the track decreases to negligibly small values. No artificial energy cut-off in the tracing of electrons and valence holes is used[25]. The entire MC procedure is averaged over 50,000 iterations to obtain reliable statistics.
The energy is transferred to the target atoms *via* three primary mechanisms: (a) the quasi-elastic scattering of electrons, (b) the quasi-elastic scattering of valence holes (coupling to phonons), and (c) atomic acceleration due to nonthermal changes in the interatomic potential induced by the electronic excitation[8,26]. The latter mechanism is approximated as a conversion of the potential energy of the electron-hole pairs into kinetic energy of the atomic lattice[8,15]. The total energy delivered to atoms is, therefore, obtained by adding the potential energy of the electron-hole pairs to the energy transferred to atoms *via* electrons and valence holes quasi-elastic scattering, simulated in TREKIS-3; this method was validated against experiments, demonstrating a reasonable agreement[7,9,15,27].

The MC-calculated radial distributions of energy density are used to set the initial conditions for atoms[28].

### SI-4.1.1 Results

### SI-4.1.1.1 Calculated energy losses and mean free paths

The single-pole approximation used for inelastic scattering cross sections of the ion, electrons and valence holes is validated by comparison of the ion stopping power with the SRIM code[29] and the electron mean free paths with the NIST database[30]. Both are in good agreement (Fig.S7 and Fig. S8, respectively), confirming the applicability of the CDF formalism and the calculated cross-sections. The differences between the beta and gamma phases of $Ga_2O_3$ are negligible, except for the low-energy electron quasi-elastic scattering.

Figure S9 shows the calculated mean free paths of the valence holes, which are also comparable between the two phases of the material. The shorter quasi-elastic mean free path at low energies in the ß phase of Ga₂O₃ may result in larger energy transfer to the lattice, but the effect is small, as discussed below and in the main text.

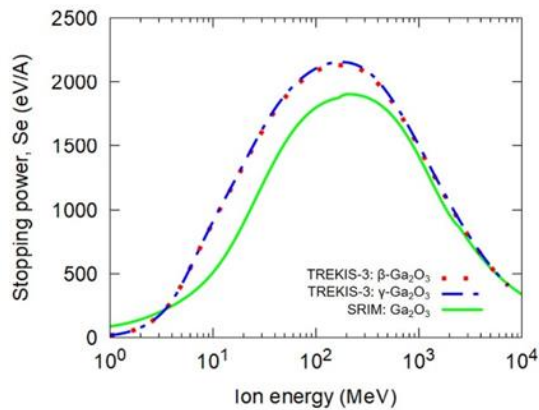

**Figure S7.** Calculated energy losses of Kr ion in β-Ga₂O₃ and γ-Ga₂O₃ compared with SRIM data[29].

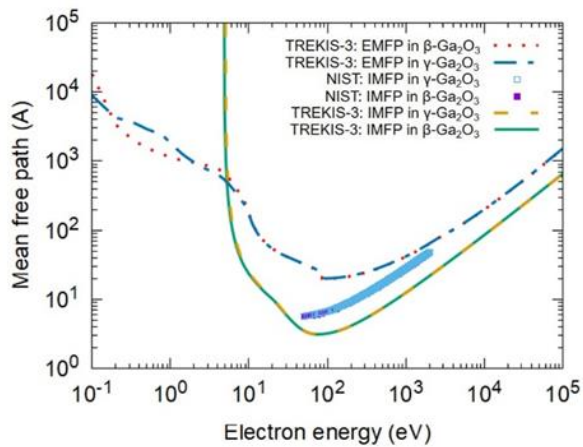

**Figure S8.** Calculated elastic (EMFP) and inelastic (IMFP) mean free paths of electrons in β-Ga₂O₃ and γ-Ga₂O₃ compared with NIST[30].

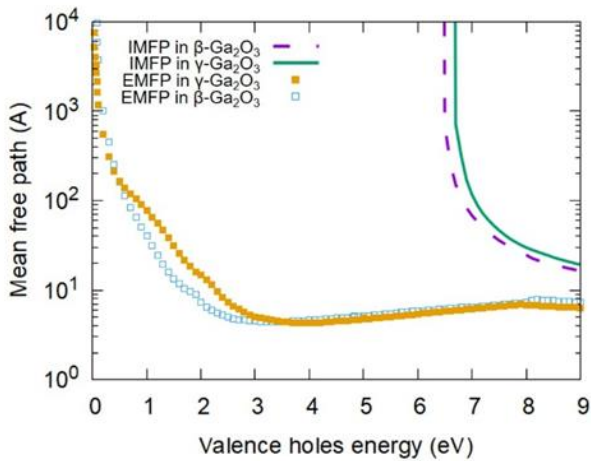

**Figure S9.** Calculated elastic (EMFP) and inelastic (IMFP) mean free paths of valence band holes in β-Ga$_2$O$_3$ and γ-Ga$_2$O$_3$.

### SI-4.1.1.2 Energy transfer to the lattice

The calculated energy losses of 147 MeV Kr in β- and γ-Ga$_2$O$_3$ are 2129 eV/Å and 2150 eV/Å, respectively. Nearly the same ion stopping power and electron scattering cross-sections results in nearly identical energy transferred to the atomic lattice in both phases of the compound, see Fig.S10. This figure shows the contribution of the three channels discussed above to the transferred energy at three-time instants after the SHI passage. Various theoretical approaches and laser-based experiments suggest that the electron–phonon coupling channel is too slow to provide the necessary energy transfer in SHI tracks[8]. Nonthermal atomic acceleration provides the required additional energy transfer before electrons cool down forming a crucial mechanism of atomic heating in swift heavy ion tracks[31]. The nonthermal atomic heating can be modelled as the transfer of the excess energy of electron–hole pairs to the atomic kinetic energy[8]. However, there is an ambiguity in the time instant at which the energy should be delivered to atoms, since the spatial distribution is evolving in time (see Fig.S10). In the case of Ga$_2$O$_3$, the best agreement with the experimental tracks is achieved when energy is transferred at 60 fs after the SHI passage. Further research is required to identify the characteristic timescales of the nonthermal effects by means of *ab initio* methods such as DFT-MD or TBMD[31].

In the γ-phase, there is only slightly smaller energy density at the radii from a few to a few tens of angstroms. In both phases, the lattice heating mainly occurs through electrons and holes quasi-elastic scattering, whereas the nonthermal process (potential energy of electron-hole pairs) contributes half as much energy to the atoms in the track center, highlighting the importance of accounting for the kinetics of valence holes when modelling the effects of SHI irradiations in materials[32–34]. The electrons and the valence holes density within 100 Å from the ion trajectory by the end of the electron cascades is also nearly the same in both phases. That indicates that the initial conditions for the heating of the atomic system are almost independent of the phase in the case of β- and γ-Ga₂O₃.

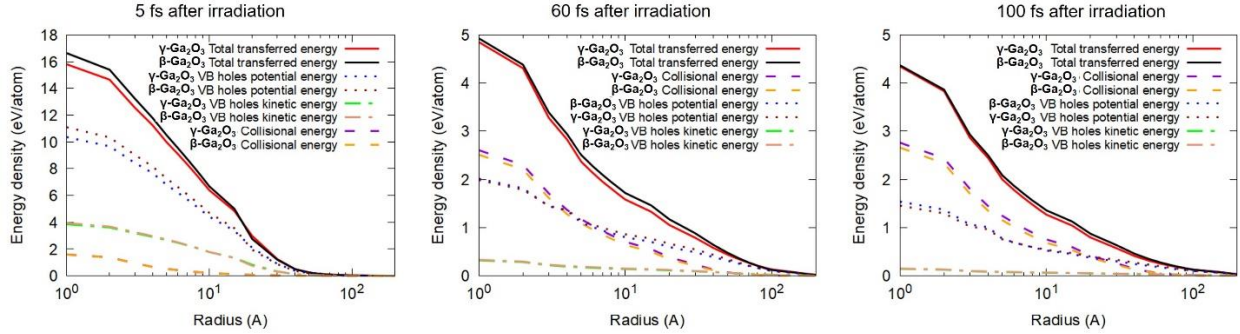

**Figure S10.** Calculated radial density of energy transferred to the atomic lattice at 5 fs, 60 fs and 100 fs after irradiation with Kr 147 MeV on β-Ga₂O₃ and γ-Ga₂O₃.

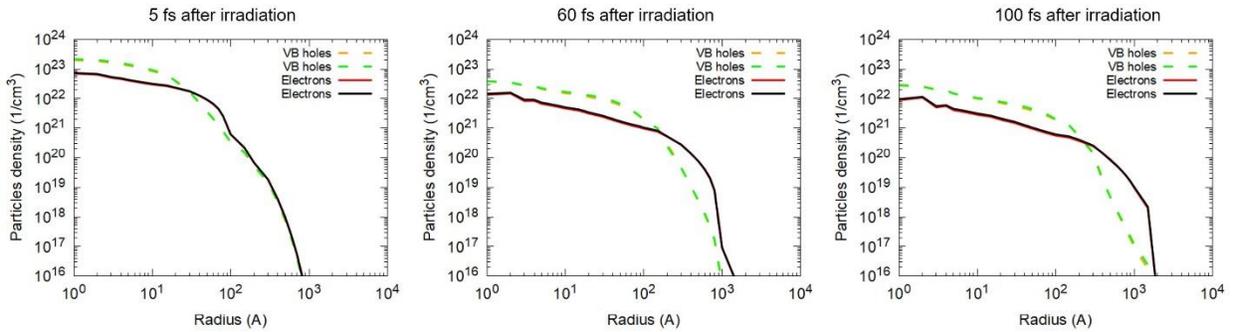

**Figure S11.** Calculated radial distribution of electron and valence band hole density in β-Ga₂O₃ and γ-Ga₂O₃ 5 fs, 60 fs and 100 fs after irradiation with 147 MeV Kr ion.

As there are experimental uncertainties about the materials bandgap[35], we performed an additional set of simulations with varying the bandgap of the material, see Fig.S12. They make a very small difference in the final energy deposition into the atomic system, since the quasi-elastic scattering is unaffected by the bandgap, and the transport of electrons and valence-band holes are only mildly influenced by the changes.

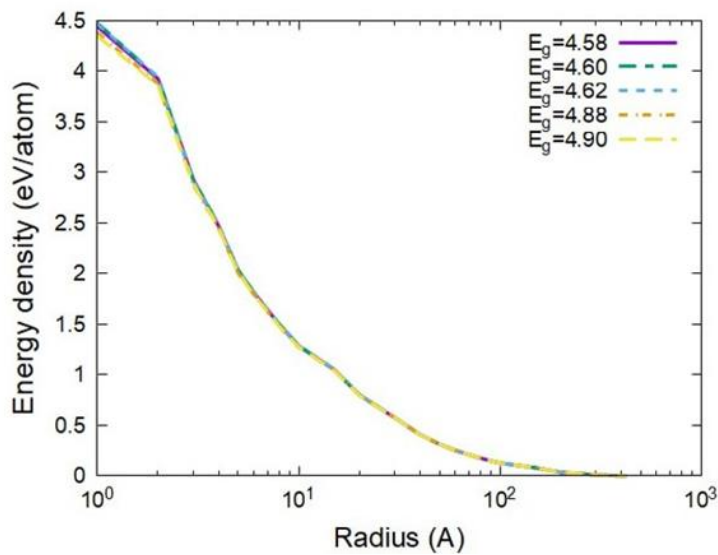

**Figure S12.** Calculated radial densities of energy transferred to the atomic lattice at 100 fs after irradiation with Kr 147 MeV for different band gap values in γ-Ga₂O₃ reported in the literature[35].

## SI-4.2. MD modelling extended

### SI-4.2.1 Energy deposition from MC simulation

The total energy, as predicted by the MC simulations (see Fig.S12), is introduced as kinetic energy imparted to the atoms in the MD simulation. This is achieved by assigning random velocities to atoms according to their distance from the ion trajectory (given by the line x, y = center of the cell) at the start of the simulation. The figure must be read as a histogram, i.e., the point plotted at 1 nm represents the value for the interval from 0 nm to 1 nm, the point at 2 nm the interval from 1 nm to 2 nm, and so on. There are many possible ways to add the kinetic energy. One could, for instance, place the extra velocity strictly parallel to the atom's existing velocity, or choose a completely random direction and sample its length from some distribution. Here we choose a direction that is uniformly distributed in three dimensions and assign it the magnitude $v_{kick} = (2 \, \Delta K / m_i)^{(1/2)}$, where $m_i$ is the mass of the atom, and $\Delta K$ the kinetic energy increment. Because the added vector is in general not perpendicular to the original velocity, the kinetic-energy change of an individual atom is not exactly $\Delta K$, but $\Delta K_{actual} = \Delta K + m v v_{kick} \cos(\alpha)$, where $\alpha$ is the angle between kick and the previous direction. However, the latter term averages to zero in each bin and therefore causes correct mean energy deposition without forcing every atom to have identical increment.

### SI-4.2.2 Annealing as accelerated long-term evolution

Annealing is often used in MD simulations to qualitatively probe the long-term evolution[36–38]. This approach sacrifices the exact time dependence of the studied process and may alter the order in which events occur[38]. Nevertheless, annealing is also employed experimentally to accelerate phase transitions[39], consistent with the principle known as the law of recrystallization, which states that recrystallization occurs faster at elevated temperatures[40]. Here, we use high-temperature annealing in MD for both the β- and γ-phase to investigate the long-term stability of the tracks.

To elucidate the recrystallization of the tracks at 300 K, we examine the evolution of disorder in the O-sublattice in both phases. Due to the timescale limitations inherent to MD simulations, we are constrained to high-temperature regimes to observe these processes. Therefore, we perform annealing simulations at 1250 K, 1200 K, 1150 K, 1100 K and 1000 K to investigate how the recrystallization rate depends on temperature.

The annealing procedure is as described in the main text: the temperature is raised in anisotropic NPT at a rate of 12 K/ps at 0 GPa and subsequently let to evolve in NVT at the target temperature. During annealing, the degree of disorder in the O-sublattice is monitored by tracking the number of O atoms in non-FCC positions using the PTM method in LAMMPS. As recrystallization progresses accelerated by the high temperature, the O atoms gradually return to FCC stacking, leading to a reduction in the non-FCC O count. Each simulation is terminated when the non-FCC O count has decreased to half its initial value at the onset of annealing. For statistical reliability, five independent simulations are conducted at each temperature.

The resulting data are presented in Fig.S13, with the MD results highlighted in the enlarged insert. By fitting an Arrhenius equation[41] to the MD data, we approximate a relation between the recrystallization rate (expressed as the half-life of the disorder) and temperature. This relation is then extrapolated to 300 K to predict the long-term evolution of the tracks at room temperature. For the β-phase, the predicted mean half-life for O-sublattice recovery at 300 K is 116 seconds, with a 95% confidence interval of [7.79e+00, 1.74e+03] seconds. While for the γ-phase, the predicted mean is 32.4 seconds, with a 95% confidence interval of [4.15e-01, 2.53e+03] seconds. For both phases, the recovery time is much shorter than the time elapsed between irradiation and STEM imaging. These simulations resulted in similar average barriers for both phases: $0.9 \pm 0.030$ eV and $0.87 \pm 0.048$ eV for the β- and γ-phase, respectively. Although the barriers are very similar, the recovery in the γ-phase proceeds slightly faster, which is expected, since this polymorph recovers back to its original structure.

These results suggest that the disorder in the O-sublattice is not permanent and the O atoms are expected to recover to the FCC structure in both phases even at room temperature. Note that this analysis does not include the Ga-sublattice, which is known to be stable at room temperature[42].

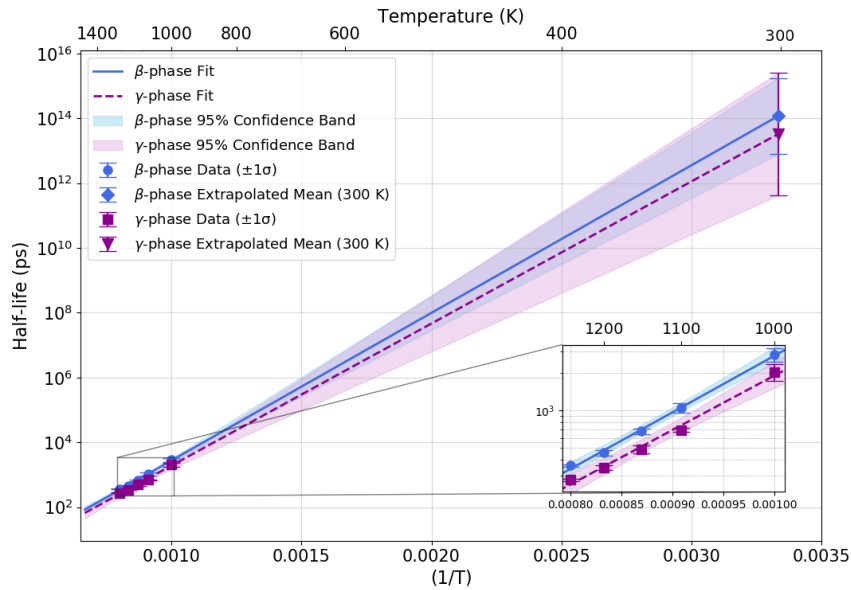

**Figure S13.** Plot showing the Arrhenius equation fit, $\tau = A\ exp(-E_a/(k_b T))$, where $\tau$ is the time required for the track size to halve, and $E_a$ and A are fitting parameters. Also shown are the extrapolation to room temperature and the 95% confidence interval. The fits apply only to the oxygen sublattice. Results for both the γ- and β-phases are presented.

## SI-4.2.3 Impact with energy transferred 5 fs after irradiation in γ-Ga₂O₃

As discussed in SI-4.1.1.2, the best agreement with experiments is obtained across both phases with the energy transferred to the lattice 60 fs after irradiation. However, due to the ambiguity in the choice of the time instant, we perform the simulations also with the energy transferred to the lattice 5 fs after irradiation in the γ-phase (see Fig.S10) to show that increasing the deposited energy and consequently the initial disorder, does not change the outcome.

In Fig.S14 the cross-section of the γ-phase simulation cell is shown at three stages after the energy has been deposited to the lattice; 5 ps after the simulated impact (a), 200 ps after the impact, which marks the end of the system's evolution in *NVE* and finally upon long-term evolution (c). In panel (a) of Fig. S14 the amorphous region depicted by the column of blue-colored O atoms in non-FCC positions has a diameter of 8.39 ± 0.19 nm, by 200 ps the disordered region has decreased to 6.45 ± 0.51 nm as recrystallization of the O-sublattice and rearranging of the Ga-atoms begins from the outer borders of the disordered track already during the impact simulation. The long-term evolution is simulated as explained in the computational methods of the main text (4 ns at 1250 K), upon which, the O-sublattice has recovered as evidenced by the color change of the O atoms in Fig. S14(c), solely leaving behind point defects as over 93% of the O-sublattice recovers.

This showcases the spectacular ability of the γ-phase to heal and suggests that the energy transferred to the system could be increased beyond the 60 fs deposition showed in the main text while still observing the recovery process of the γ-phase returning to its original structure.

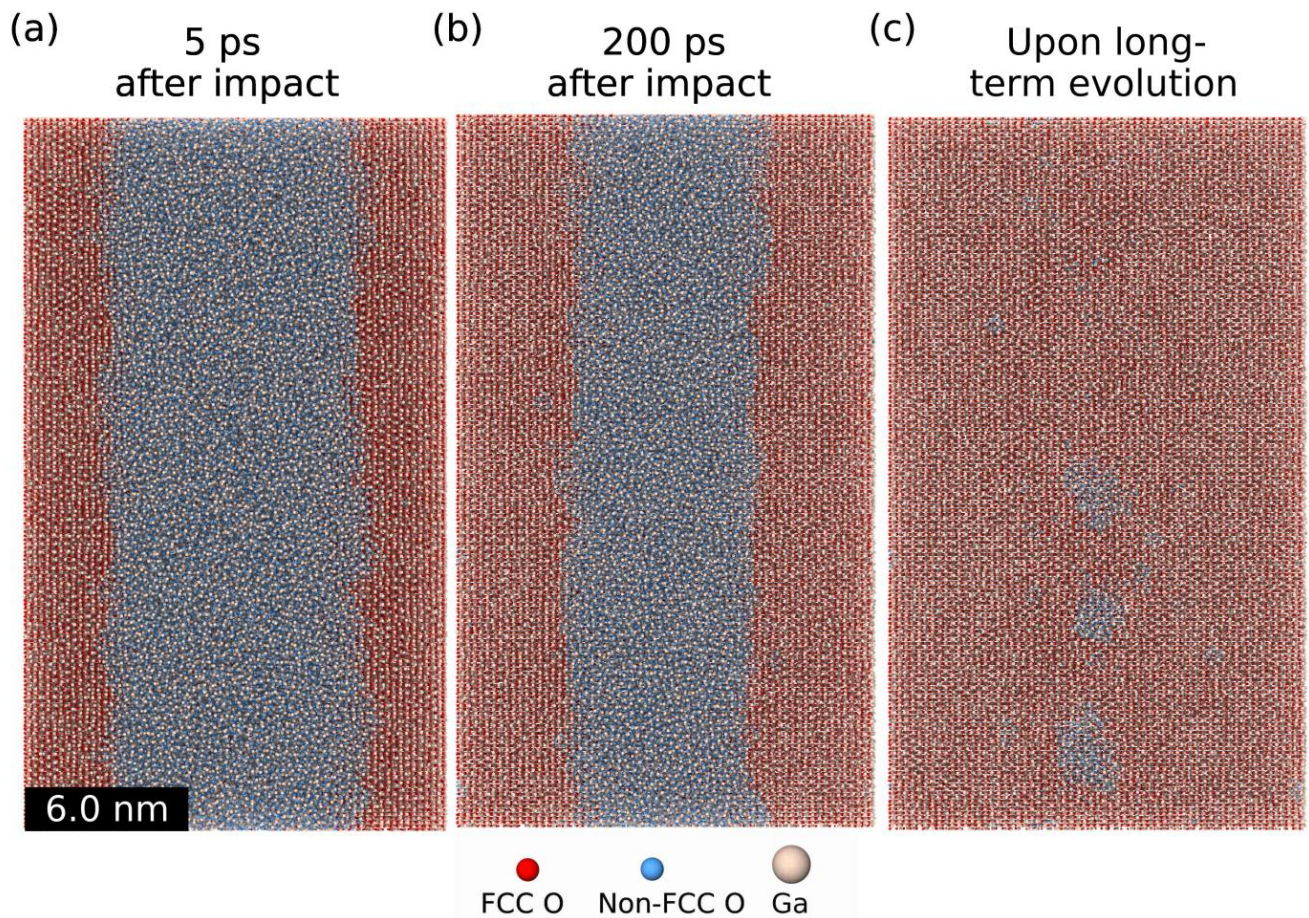

**Figure S14.** Cross-section of the γ-phase simulation cell (a) 5 ps and (b) 200 ps after impact with energy deposition of energy transferred to the lattice 5 fs after irradiation and (c) upon long-term evolution. Gallium atoms are beige while oxygen atoms in FCC positions are red, and in non-FCC positions are blue.

**Supplementary note 5: Additional data of back-loop STEM simulations**

As discussed in the main text in connection to the interpretations of Fig.4, there is a good match between the back-looped simulated STEM image of the tracks and the experimental data. For clarity, Fig.S15 shows a side-by-side comparison of the simulated HAADF image (a) taken from Fig.4 with an experimental HAADF STEM image (b) of an ion track from the 147 MeV Kr irradiated $\beta$-$Ga_2O_3$. As mentioned earlier, the sizes of $\gamma$-tracks in $\beta$-$Ga_2O_3$ matrix may slightly vary and deviate from the exact size obtained in MD simulations; likely because of the stochastic lateral distribution of tracks implying different proximity to the nearby tracks and as such different conditions for the overlap or other forms of interactions. Still, there is excellent correspondence between simulated and experimental images in terms of atomic column arrangement inside and outside the tracks. Notably, yellow arrows indicate distorted Ga columns at the track periphery present in both simulation and experiment. Through the transition from the monoclinic to cubic phase, an intermediate structure is formed at the track periphery where Ga columns appear in dumbbell-like pairs. Figure S15(c) shows a simulated HAADF STEM image of $\gamma$-$Ga_2O_3$ (see Fig.4(d)) next to an enlarged view of the track core regions indicated in Fig.S15(a) and (b), in (d) and (e) respectively. Good agreement between all three images is obvious.

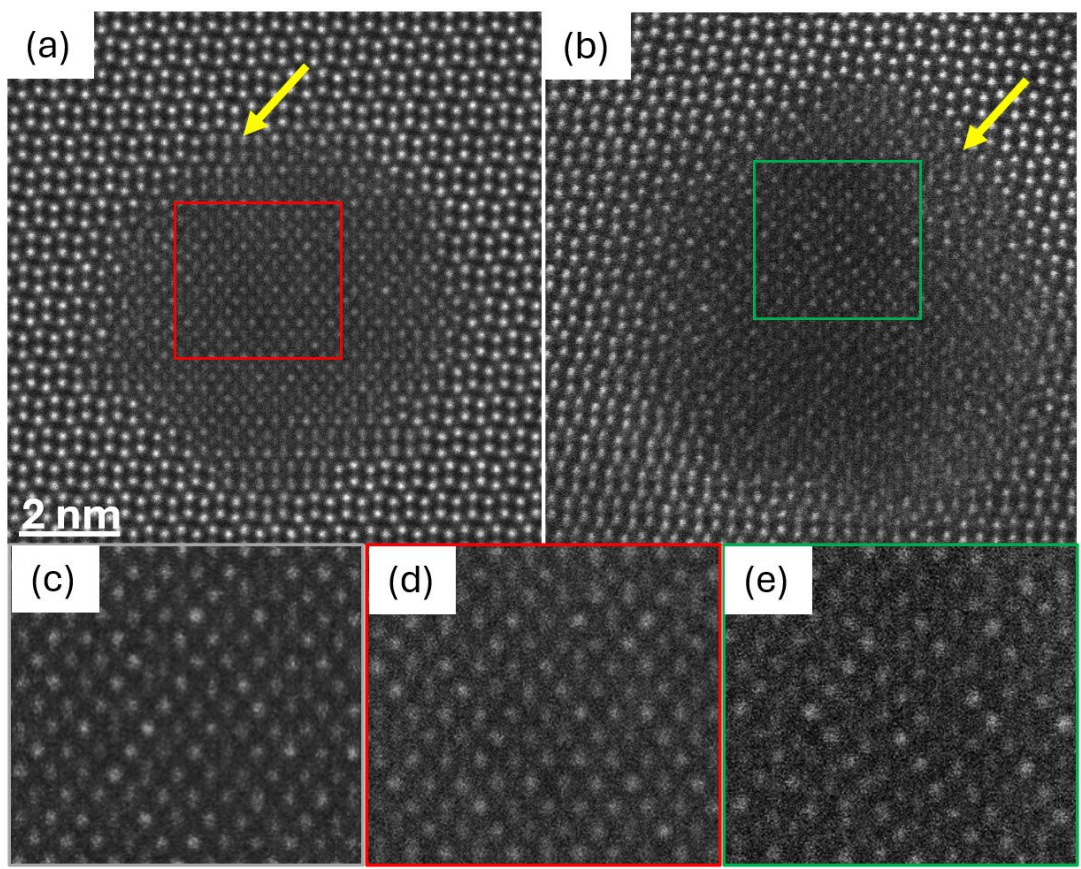

**Figure S15.** Side-by-side comparison of the simulated HAADF image in Fig.4(d) and experimental HAADF STEM image of a 147 MeV Kr tracks in $\beta$-$Ga_2O_3$.